\documentclass[twocolumn,trackchanges]{aastex631}

\usepackage{CJK}

\newcommand{\petit}{\texttt{petitRADTRANS}}

\newcommand{\dynesty}{\texttt{dynesty}}
\newcommand{\teff}{T$_{\rm eff}$}

\newcommand{\kms}{$\rm km\,s^{-1}$}
\newcommand{\keltb}{KELT-20~b}
\newcommand{\caltech}{Department of Astronomy, California Institute of Technology, Pasadena, CA 91125, USA}
\newcommand{\gps}{Division of Geological \& Planetary Sciences, California Institute of Technology, Pasadena, CA 91125, USA}
\newcommand{\ucsc}{Department of Astronomy \& Astrophysics, University of California, Santa Cruz, CA95064, USA}
\newcommand{\keck}{W. M. Keck Observatory, 65-1120 Mamalahoa Hwy, Kamuela, HI 96743, USA}
\newcommand{\ucla}{Department of Physics \& Astronomy, 430 Portola Plaza, University of California, Los Angeles, CA 90095, USA}
\newcommand{\jpl}{Jet Propulsion Laboratory, California Institute of Technology, 4800 Oak Grove Dr.,Pasadena, CA 91109, USA}
\newcommand{\ucsd}{Center for Astrophysics and Space Sciences, University of California, San Diego, La Jolla, CA 92093}

\newcommand{\northwestern}{Center for Interdisciplinary Exploration and Research in Astrophysics (CIERA) and Department of Physics and Astronomy,
Northwestern University, Evanston, IL 60208, USA}
\newcommand{\arizona}{James C. Wyant College of Optical Sciences, University of Arizona,
Meinel Building 1630 E. University Blvd., Tucson, AZ 85721, USA}

\received{\today}
\shorttitle{KPIC emission spectroscopy of KELT-20~b}
\shortauthors{Finnerty et al.}
\graphicspath{{./}{}}

\begin{document}
\begin{CJK*}{UTF8}{gbsn}

\title{Water dissociation and rotational broadening in the atmosphere of KELT-20 b from high-resolution spectroscopy}

\correspondingauthor{Luke Finnerty}
\email{lfinnerty@astro.ucla.edu}

\author[0000-0002-1392-0768]{Luke Finnerty}
\affiliation{\ucla}

\author[0000-0002-6171-9081]{Yinzi Xin}
\affiliation{\caltech}

\author[0000-0002-6618-1137]{Jerry W. Xuan}
\affiliation{\caltech}

\author{Julie Inglis}
\affiliation{\caltech}

\author[0000-0002-0176-8973]{Michael P. Fitzgerald}
\affiliation{\ucla}

\author[0000-0003-2429-5811]{Shubh Agrawal}
\affiliation{Department of Physics and Astronomy, University of Pennsylvania, Philadelphia, PA 19104, USA}

\author[0000-0002-6525-7013]{Ashley Baker}
\affiliation{\caltech}

\author{Randall Bartos}
\affiliation{\jpl}

\author{Geoffrey A. Blake}
\affiliation{\gps}


\author[0000-0003-4737-5486]{Benjamin Calvin}
\affiliation{\caltech}
\affiliation{\ucla}

\author{Sylvain Cetre}
\affiliation{\keck}

\author[0000-0001-8953-1008]{Jacques-Robert Delorme}
\affiliation{\keck}
\affiliation{\caltech}

\author{Greg Doppmann}
\affiliation{\keck}

\author[0000-0002-1583-2040]{Daniel Echeverri}
\affiliation{\caltech}

\author[0000-0001-9708-8667]{Katelyn Horstman}
\affiliation{\caltech}
\altaffiliation{NSF Graduate Research Fellow}

\author[0000-0002-5370-7494]{Chih-Chun Hsu}
\affiliation{\northwestern}

\author[0000-0001-5213-6207]{Nemanja Jovanovic}
\affiliation{\caltech}

\author[0000-0002-4934-3042]{Joshua Liberman}
\affiliation{\caltech}
\affiliation{\arizona}

\author[0000-0002-2019-4995]{Ronald A. L\'opez}
\affiliation{\ucla}

\author[0000-0002-0618-5128]{Emily C. Martin}
\affiliation{\ucsc}

\author{Dimitri Mawet}
\affiliation{\caltech}
\affiliation{\jpl}

\author{Evan Morris}
\affiliation{\ucsc}

\author{Jacklyn Pezzato}
\affiliation{\caltech}

\author[0000-0003-2233-4821]{Jean-Baptiste Ruffio}
\affiliation{\ucsd}

\author[0000-0003-1399-3593]{Ben Sappey}
\affiliation{\ucsd}

\author{Tobias Schofield}
\affiliation{\caltech}

\author{Andrew Skemer}
\affiliation{\ucsc}

\author{Taylor Venenciano}
\affiliation{Physics and Astronomy Department, Pomona College, 333 N. College Way, Claremont, CA 91711, USA}

\author[0000-0001-5299-6899]{J. Kent Wallace}
\affiliation{\jpl}

\author[0000-0003-0354-0187]{Nicole L. Wallack}
\affiliation{Earth and Planets Laboratory, Carnegie Institution for Science, Washington, DC 20015, USA}

\author[0000-0003-0774-6502]{Jason J. Wang (王劲飞)}
\affiliation{\northwestern}

\author[0000-0002-4361-8885]{Ji Wang (王吉)}
\affiliation{Department of Astronomy, The Ohio State University, 100 W 18th Ave, Columbus, OH 43210 USA}



\begin{abstract}
We present atmospheric retrievals from Keck/KPIC phase II observations of the ultra-hot Jupiter KELT-20/MASCARA-2~b. Previous free retrievals of molecular abundances for ultra-hot Jupiters have been impacted by significant model biases due to variations in vertical abundance profiles, which we address by including molecular dissociation into our retrieval framework as an additional free parameter. We measure the abundance of CO ($\rm \log CO_{MMR} = -2.5^{+0.6}_{-0.5}$) and obtain a lower limit on the abundance of H$_2$O ($\rm \log H{_2}O_{MMR} = -1.5^{+0.8}_{-1.0}$, $>-3.0$ at 95\% confidence) in the atmosphere of \keltb. These abundances yield an atmospheric $\rm C/O = 0.1^{+0.4}_{-0.1}$ ($\rm C/O < 0.9$ at 95\% confidence)  and suggest a metallicity approximately solar to $10\times$ solar. H$_2$O is dissociated at pressures below $\log P_{\rm H_2O} = -1.2^{+0.5}_{-0.7}$ bar, roughly consistent with predictions from chemical equilibrium models, and suggesting that the retrieved composition is not a result of assumptions about the vertical mixing profiles. We also constrain the rotational velocity of \keltb\ to $v\sin i = 7.5\pm0.7$ \kms, suggesting the presence of a jet comparable to the sound speed in the direction of the planet's rotation, assuming the actual rotation of the planet is tidally locked. 
\end{abstract}

\keywords{Exoplanet atmospheres (487) --- Exoplanet atmospheric composition (2021) --- Hot Jupiters (753) --- High resolution spectroscopy (2096)}


\section{Introduction} \label{sec:intro}

 \begin{deluxetable}{ccc}
    \tablehead{\colhead{Property} & \colhead{Value} & \colhead{Ref.}}
    \startdata
        & \textbf{KELT-20} & \\
        \hline
        RA & 19:38:38.7 &  \citet{gaiaedr3} \\
        Dec & +31:13:09 &  \citet{gaiaedr3} \\
        Sp. Type & A2V & \citet{lund2017} \\
        $K_{\rm mag}$ & $7.415\pm0.017$ & \citet{cutri2003}  \\
        Mass & $1.76^{+0.14}_{-0.19} M_\odot$ & \citet{lund2017}  \\
        Radius & $1.60\pm0.06 R_\odot$ & \citet{talens2018} \\
        \teff & $8730^{+250}_{-260}$ K & \citet{lund2017}  \\
        $\log g$ & $4.29^{+0.017}_{-0.020}$ &  \citet{lund2017} \\
        $v\sin i$ & $114\pm3$ \kms & \citet{talens2018} \\
        $v_{\rm rad}$ & $-23.3\pm0.3$ \kms & \citet{lund2017}  \\
        $\rm [Fe/H]$ & $-0.02\pm0.07$  & \citet{talens2018}   \\
        \smallskip \\
        \hline
         & \textbf{KELT-20 b} & \\
        \hline
        Period [day] &  $3.4741012\pm2\times10^{-7}$ & \citet{exofop3} \\
        $\rm t_{\rm transit}$ [JD] & $2460337.98638\pm5\times10^{-5}$ & \citet{exofop3}  \\
        $a$ & $0.057\pm0.006$ AU &  \citet{talens2018}  \\
        $i$ & $86.4^{\circ+0.5}_{-0.4}$ & \citet{talens2018}  \\
        $K_{\rm p}$ & $176.7\pm0.6$ \kms & \citet{yan2022}  \\
        Mass & $<3.372 M_J$ & \citet{lund2017} \\
        Radius & $1.83\pm0.07$ $R_J$ & \citet{talens2018} \\
        $\rm T_{\rm eq}$ & $2262\pm73$ K & \citet{talens2018}  \\
        C/O & $<0.9$ & This work  \\
        $\rm [C/H]$ & $0.1^{+0.6}_{-0.5}$ &  This work \\
        $\rm [O/H]$ & $1.1\pm0.8 (>-0.1)$ &  This work \\
        $\rm [(C+O)/H]$ & $ 1.0\pm0.7 (>-0.15)$ & This work \\
        $v\sin i$ & $7.5\pm0.7 $ \kms & This work
    \enddata
    \caption{Stellar and planetary properties for the KELT-20 system.} 
    \label{tab:props}
\end{deluxetable}

Ultra-hot Jupiters (UHJs), hot Jupiters with equilibrium temperatures in excess of 2000 K, are an ideal exoplanet population for atmospheric characterization studies. The extreme conditions of these planets and complicated dynamical history suggested by the close-in, misaligned orbits of many UHJs \citep{naoz2011, rice2022} makes these planets compelling targets for efforts to improve our observational understanding of atmospheric circulation and planet formation. 

However, molecular abundance estimation for UHJs is complicated by the impacts of thermal dissociation, vertical mixing, cold trapping, and photochemistry. While CO only dissociates at very high temperatures and low pressures, typical UHJ conditions are expected to dissociate H$_2$O at pressures deeper than $\sim1$ mbar under chemical equilibrium. If abundances are vertically fixed in a retrieval, the only parameter which can reduce the strength of H$_2$O features to match the effects of dissociation is the H$_2$O abundance  itself. As a result, retrievals which assume fixed abundances with altitude will be biased towards high-C/O, high-metallicity compositions \citep[e.g.][]{brogi2023, ramkumar2023}.

While assuming chemical equilibrium abundance profiles can resolve this issue \citep{brogi2023, ramkumar2023}, local chemical equilibrium based on an assumed $P-T$ profile is a potentially restrictive assumption. Internal heating and the resulting vertical mixing can lead to severe depletion of species expected from models which use only the local temperature to calculate equilibrium abundances \citep{sing2024, welbanks2024}. Cold trapping has similarly been found to cause significant depletion of condensable species on planets with strong day/night temperature contrasts, which is not reproducible with 1D local chemical equilibrium models \citep{pelletier2023, hoeijmakers2024}. As a result of these effects, local chemical equilibrium retrievals may be pushed to inaccurate abundances in order to minimize the impact of species present under the local chemical equilibrium assumption but absent in the real atmosphere. Free retrievals can adjust to the presence or absence of species based on these effects without biasing the abundances of other species.

UHJs around A-type stars are also exposed to extreme UV fluxes, which may drive photochemical processes in the upper atmosphere which will not be captured by a thermochemical equilibrium model. Photochemically produced species have been detected in \textit{JWST} observations of hot/warm Jupiters \citep{tsai2023, dyrek2024}, providing an observational example of the limits of an equilibrium assumption. While there are publicly available tools to estimate abundances which include photochemistry effects \citep{vulcan}, these tools are too slow to use for retrievals, and the reaction networks used in these models need to be observationally verified. This creates a need for free retrieval frameworks which can account for vertically-varying molecular abundances. 

UHJs also provide an opportunity to directly test Global Circulation Model (GCM) predictions in extreme conditions, provided the planetary ephemeris and spectrograph line-spread function (LSF) are sufficiently well characterized to robustly measure velocity offsets and line broadening. Under UHJ conditions, GCMs predict supersonic day-to-night winds up to 10 \kms \citep{wardenier2021, komacek2022, beltz2022}. While medium/low resolution spectroscopy cannot measure these supersonic wind speeds directly, high-resolution observations have been able to provide some wind measurements consistent with these models \citep[e.g.][]{pai2022, gandhi2023}. For tidally locked planets such as UHJs, constraining the rotational velocity in emission can provide an independent measurement of the equatorial jet speed to complement velocity shift measurements from high-resolution spectroscopy obtained during transit \citep{lesjack2023, finnerty2024, bazinet2024}. Interpreting shifts measured during transit as wind speeds has been difficult due to the potential for chemical differences between leading and trailing limbs \citep{wardenier2021, savel2022}. Combining measurements of jet velocities from multiple observing geometries is therefore critical to understanding atmospheric circulation for hot Jupiters and provides a unique observational test of GCMs in an extreme regime \citep{pino2022, vansluijs2023}.

We observed \keltb\ (also known as MASCARA-2~b) as part of our ongoing Keck/KPIC hot Jupiter survey \citep{finnerty2023, finnerty2023spie, finnerty2024, finnerty2025}. Properties of the system are summarized in Table \ref{tab:props}. First discovered in 2017 \citep{lund2017}, \keltb\ is in many ways a typical UHJ, with an equilibrium temperature of 2260 K, an inflated radius \citep[$1.83\pm0.07R_J$][]{talens2018}, and a rapidly rotating \citep[$v\sin i = 114$ \kms][]{talens2018} early-type host star. However, compared with other UHJs, \keltb\ has a longer orbital period of 3.5 days, compared with the $<2$ day periods of e.g. WASP-33~b, KELT-9~b, MASCARA~1b, and TOI-1518~b. Spin-orbit misalignment measurements using the Rossiter-McLaughlin Effect have also found that the orbital plane \keltb\ is well-aligned with the host star's rotation \citep{talens2018}, in contrast with the other UHJs listed previously \citep{collier2010, gaudi2017, talens2018m1, cabot2021}. This suggests \keltb\ could have formed and/or migrated via a different mechanism than other UHJs, perhaps without the late high-eccentricity migration that is believed to be necessary to produce retrograde UHJs such as WASP-33~b and KELT-9~b \citep{naoz2011}. 

\keltb\ has been extensively studied at optical wavelengths in both emission and transmission \citep[see][and references therein]{johnson2023, petz2024}, revealing a variety of transition metal species but no evidence for TiO, VO, or CaH in the optical. In the infrared, strong H$_2$O and CO emission features were reported from low-resolution \textit{HST} and \textit{Spitzer} spectroscopy \citep{fu2022}. Subsequently, \citet{kasper2023} reported the results of a joint retrieval of these data and MAROON-X optical emission data which suggested a super-solar metallicity and C/O ratio. In this paper, we present retrievals based on Keck/KPIC $K$-band emission observations (2.1--2.48$\mu$m), providing the first high-spectral-resolution detections of CO and H$_2$O in \keltb\ and a constraint on the atmospheric C/O ratio from these gases. 

We discuss the KPIC observations of \keltb, data reduction procedure, and retrieval setup in Section \ref{sec:obs}. Retrieval results are presented in Section \ref{sec:res}. We discuss these results and the implications for the formation history and atmospheric dynamics of \keltb\ in Section \ref{sec:disc}, and summarize our results in Section \ref{sec:conc}.

\section{Observations, Data Reduction, and Retrieval}\label{sec:obs}

\subsection{Observations}

The KELT-20 system was observed with Keck II/KPIC phase II \citep{nirspec, nirspecupgrade, nirspecupgrade2, kpic, echeverri2022, jovanovic2025} on UT 30 July 2023 from 5:31 to 12:06 UTC, totaling just over 6.5 hours. The observations began shortly after the end of the \keltb\ secondary eclipse, and covered a phase range of 0.539--0.617. During this time, the nominal star-frame planet velocity went from $-43.2$ \kms\ to $-119.5$ \kms. The exposure time was 300 seconds per frame, which kept the velocity shift within a single exposure to $<1.1$ \kms, substantially smaller than the $\sim 9$ \kms\ resolution of NIRSPEC \citep{nirspecupgrade2}, in order to minimize smearing of the planet signal within a single exposure. Conditions and instrument performance were typical for KPIC phase II, with a top-of-atmosphere throughput of 3-4\% throughout the observing sequence. The reduced/extracted spectra from each exposure had a signal-to-noise ratio (SNR) of approximately 120-140 per pixel in the orders used for retrieval. 

\keltb\ was observed using only a single KPIC science fiber, rather than nodding between fibers as in \citet{finnerty2023, finnerty2024}. This reduced overheads associated with switching fibers and allowed all observations to be taken using the highest-throughput fiber to maximize total signal-to-noise. This approach also simplified handling of the KPIC line-spread function (LSF), which is slightly different for each fiber. KPIC has minimal background in the $K$-band, which is sufficiently removed using dark frames taken before/after observing to not impact retrievals. 

\subsection{Data Reduction}

A late-type star (HIP 95771, M0.5IIIb) was observed to provide wavelength calibration using the KPIC Data Reduction Pipeline (DRP)\footnote{\href{https://github.com/kpicteam/kpic_pipeline/}{https://github.com/kpicteam/kpic\_pipeline/}}. As with previous KPIC $K$-band observations, three orders (37--39, 1.94--2.09$\mu$m) are severely impacted by telluric CO$_2$ and are omitted from the analysis. While previous KPIC observations struggled with wavelength calibration in orders 35 and 36 due to an absence of strong stellar or telluric lines, in this case the wavelength calibrator was observed at a median airmass of 1.32, leading to somewhat deeper telluric lines and a wavelength solution which appears to be reliable in all orders. We therefore include orders 31--36 in the analysis. The improved calibration in the bluer orders may be a result of iterative improvements in the KPIC wavelength calibration from using previous wavelength solutions as the starting point for optimizing a new wavelength solution. 

The KPIC DRP was used to produce background and bad pixel frames for the 300s exposure time. Ten background frames were obtained in both the afternoon prior to observing and the morning after observing, enabling a high-quality background subtraction even without nodding between the KPIC fibers.

The flux extraction differed significantly from the default KPIC DRP in order to accurately handle the non-Gaussian and variable KPIC LSF. As part of the trace finding step, we also fit the trace profile in the  spatial (detector $y$) direction using a Gaussian-Hermite model, similar to the approach described by \citet{holtzman2018}. This model multiplies a Gaussian profile of variable width by the Hermite polynomials of order $\leq4$, with the amplitude of each polynomial as a free parameter. This model allow fitting of both non-Gaussian wings and asymmetries in the line profile. The raw fit parameter values are then smoothed by a 13-pixel Gaussian kernel to reduce noise and ensure variation in the LSF matches the spatial scales for such variation found in \citet{nirspecupgrade2}. 

This spatial profile is then used to compute weights for optimal extraction \citep{horne1986} and extract 1D spectra. To estimate the KPIC LSF, we multiply the overall width parameter by a factor of 1.14 in order to match a known asymmetry induced by the NIRSPEC optics \citep{Finnerty2022}. These scaled profiles are then convolved with the forward-model of the planet atmosphere during the retrieval to replicate the instrumental broadening of NIRSPEC. 

\subsection{Atmospheric Retrieval}

Our atmospheric retrieval framework has previously been described in \citet{finnerty2023, finnerty2024, finnerty2025}. We provide a brief summary here, with an emphasis on new features. The full list of parameters and priors is included in Table \ref{tab:priors}. We use a PHOENIX \citep{phoenix} model for KELT-20~A with $T_{\rm eff} = 9000$ K, $\log g = 4.5$, and solar metallicity, roughly matching the stellar parameters reported in \citet{lund2017}. The stellar model is rotationally broadened using the wavelength-dependent technique described in \citet{carvalho2023} to 114 \kms\ \citep{talens2018}. 

We use \petit 3\ \citep{prt:2019, prt:2020, Nasedkin2024} to compute model planet spectra. The mass of \keltb\ is poorly constrained, with \citet{lund2017} placing a $3\sigma$ upper limit of $M_{pl} < 3.5 M_J$, which motivates our inclusion of $\log g$ as a free parameer in the atmospheric retrieval. As in our previous works, we use 80 log-uniform spaced pressure layers from $10^2$ to $10^{-6}$ bar. Following \citet{finnerty2024}, we use the pressure-temperature ($P-T$) profile model from \citet{guillot2010}. While the intrinsic temperature parameter $T_{\rm int}$ has been fixed in other high-resolution studies \citep[e.g.][]{gibson2022, ramkumar2023}, we include it as a free parameter in our retrievals. $T_{\rm int}$ determines the lower-atmosphere behavior of the $P-T$ profile, below the thermal inversion and the bulk of the emission contribution function, but we find that its inclusion is important to breaking degeneracies between abundances and dissociation pressures.

For molecular abundances, we use a free-retrieval approach which allows for variations in vertical abundance profiles for certain species. In addition to the abundance, our model can optionally fit a dissociation pressure and a formation pressure. At pressures lower than the dissociation pressure, the abundance decays as a power law, $\left(P/P_0\right)^4$. Similarly, at pressures greater than the formation pressure, the abundance decays as $\left(P/P_0\right)^{-4}$. Each of these parameters is optional, and different parameterizations can be used for different species, including freely fitting the power law index. The default power law index of 4 was an arbitrary choice made to roughly match the profiles seen in \texttt{easyCHEM} \citep{molliere2017, lei2024} results for a nominal \keltb\ $P-T$ profile.

Based on chemical equilibrium models, we allow CO to dissociate only at low pressures, extending below the top-of-atmosphere cutoff pressure to allow for the no-dissociation scenario. For H$_2$O, we expect dissociation to occur near the thermal inversion, so we fit for the dissociation pressure. We expect OH to be produced around the same pressure as H$_2$O dissociation, so we fit for the OH formation pressure, and we expect OH will itself be dissociated at high altitudes, so we also fit for the OH dissociation pressure. C$_2$H$_2$, if present, is expected to dissociate at intermediate pressures, so we fit for the dissociation pressure. We do not fit the power-law index for any species, after initial tests found this parameter was unconstrained in retrievals.

We also fit for the overall hydrogen mass fraction and the H$_2$ dissociation pressure. As these species have only weak spectral features in the region probed, our sensitivity to these parameters is primarily indirect via collisionally-induced-absorption (CIA) opacity and the impact on the mean molecular weight. The remaining mass of the atmosphere is assumed to be helium.

The opacity tables used for H$_2$O, OH, $^{12}$CO, and $^{13}$CO were previously described in \citet{finnerty2023}. For H$_2$O we used the \citep{polyansky2018} partition function and HITEMP 2010 linelist \citep{hitemp2010}. For both CO isotopologues we used the HITEMP 2019 lists \citep{hitemp2020} and the \citet{li2015} partition function. For OH we used the partition function from \citet{YOUSEFI2018} and HITEMP 2020 linelist \citep{hitemp2020}. Opacities were generated via ExoCross \citep{exocross2018} following instructions in the \petit~documentation. For C$_2$H$_2$, we converted the aCeTeY \citep{chubb2020} opacity table available from DACE \citep{grimm2021} to the \petit~format.  As clouds are not expected under dayside UHJ conditions, we do not include any cloud species, but retain the scattering mode in \petit. See \citet{finnerty2023} for a more detailed discussion of this choice in UHJs. A test retrieval including a grey cloud deck preferred a cloud-top pressure greater than the base of the emission contribution function, indicating a preference for a clear atmosphere. H$^-$ is similarly not included, following \citet{finnerty2023}, who found that H$^-$ introduces a uniform reduction in the line strengths for small bandpass observations. In this scenario, the impact of H$^-$ becomes degenerate with changes in the scale factor and $P-T$ profile. This is not the case for analyses covering a large bandpass, such as joint optical+IR retrievals, which will likely need to include and constrain H$^-$ explicitly. 

Our retrieval pipeline uses principal component analysis (PCA) to remove time-varying instrumental fringing and telluric effects which would otherwise dominate over the planet signal, as previously described in \citet{finnerty2024}. The number of principal components to drop is difficult to choose in a robustly motivated way \citep{cheverall2023}. We therefore ran retrievals omitting 4, 6, 8 and 12 PCs in order to assess the impact of this choice on the retrieval results.

HRCCS retrievals also use an overall scaling parameter applied to the planet spectrum \cite{brogi2019}. While \citet{finnerty2023, finnerty2024} used a linear scale factor, for this work we follow \citet{finnerty2025} and switch to a truncated log-normal scale factor with a mean of 0 (corresponding to a $1\times$ multiplication of the planet spectrum) and standard deviation of 0.1. This distribution is truncated at 7.5 standard deviations from the mean for numeric stability. This is more in line with other analyses, for which a logarithmic prior for the scale factor \citep[e.g.][]{line2021} is common. Using a normal prior, rather than uniform, may help to break the temperature/scale factor degeneracy discussed in \citet{brogi2023, finnerty2024}.

The nested sampling for the retrieval was performed using \texttt{MultiNEST} \citep{feroz2008, feroz2009, buchner2014, feroz2019} with 800 live points and a $\Delta \log z = 0.01$ stopping criteria. We continue to use the \citet{brogi2019} $\log L$ function as in \citet{finnerty2023, finnerty2024}. Compared with previous retrievals using \dynesty, \texttt{MultiNest} retreivals run $\sim10\times$ faster, although the wider wavelength coverage and strict convergence used in this analysis still requires approximately one week of computing time with 16 Intel E5-2670 CPU cores. 

\section{Results}\label{sec:res}
\begin{deluxetable*}{ccccc}
    \tablehead{\colhead{Name} & Symbol  & \colhead{Prior} & \colhead{Retrieved Max-L} & \colhead{Retrieved Median}}
    \startdata
        log infrared opacity [$\rm cm^{2} g^{-1}$] & $\log \kappa$ &  Uniform(-4, 0) & -2.8 & $-2.6^{+0.8}_{-0.8}$ \\ 
        log infrared/optical opacity & $\log \gamma$  &  Uniform(0, 3) & 1.5 & $1.6^{+0.2}_{-0.2}$ \\
        Intrinsic Temperature [K] & $\rm T_{int}$ & Uniform(500,1500) & 600.0 & $840.0^{+290.0}_{-220.0} *$ \\
        Equilibrium temperature [K] & $\rm T_{equ}$ & Uniform(1500, 5000) & 1580 & $1660^{+220}_{-120.0}* (<2100)$ \\
        Surface gravity [$\rm \log cm\ s^{-2}$] & $\log g$ & Uniform(2,4.5) & 3.6 & $3.3^{+0.7}_{-0.8}*$ \\
        $K_{\rm p}$  [\kms] & $ K_{\rm p}$  & Uniform(157,197) & 177.6 & $178.5^{+3.4}_{-3.3}$ \\
        $v_{\rm sys}$ offset [\kms] & $\Delta v_{\rm sys}$ & Uniform(-20, 20) & -0.2 & $0.3^{+1.6}_{-1.5}$ \\
        Rotational velocity [\kms] & $v_{\rm rot}$ & Uniform(0, 15) & 7.1 & $7.5^{+0.7}_{-0.7}$ \\
        log H$_2$O mass-mixing ratio & log H$_2$O  &  Uniform(-12, -0.3) & -1.2 & $-1.5^{+0.8}_{-1.0} * (>-3.0)$  \\
        log CO mass-mixing ratio & log CO &  Uniform(-12, -0.3) &  -2.9 & $-2.5^{+0.6}_{-0.5}$ \\
        log OH mass-mixing ratio & log OH  & Uniform(-12, -0.5) &  -2.3 & $-2.9^{+1.1}_{-4.9}*$ \\
        log C$_2$H$_2$ mass-mixing ratio & log C$_2$H$_2$ & Uniform(-12, -1) & -4.0 & $-7.3^{+3.0}_{-3.0}*$ \\
        log Fe mass-mixing ratio & log Fe  & Uniform(-12, -1) & -2.1 & $-5.1^{+2.7}_{-4.3}*$ \\
        CO dissociation pressure & $\log \rm P_{CO}$ & Uniform(-7, -2) &  -4.8 & $-4.8^{+1.5}_{-1.4}*$ \\
        H$_2$O dissociation pressure & $\log \rm P_{H_2O}$ & Uniform(-6, 2) &  -0.5 & $-1.2^{+0.5}_{-0.7}$ \\
        OH dissociation pressure & $\log \rm P_{OH}$ & Uniform(-6, 2) & -5.3 & $-3.4^{+2.7}_{-1.7}*$ \\
        C$_2$H$_2$ dissociation pressure & $\log \rm P_{C_2H_2}$ & Uniform(-6, 2) & -0.5 & $-1.7^{+2.4}_{-2.7}*$ \\
        OH formation pressure & $\log \rm P_{form OH}$ & Uniform(-6, 2) & -2.2 & $-1.4^{+2.1}_{-2.2}*$ \\
        log $\rm ^{13}CO/^{12}CO$ & $\rm \log ^{13}CO_{rat}$ &  Uniform(-6, -0.5) &  -1.9 & $-4.8^{+2.0}_{-2.0}* (<-2.0)$ \\
        log H mass-mixing ratio & $\log \rm all H$ &  Uniform(-0.4, -0.05) & -0.1 & $-0.2^{+0.1}_{-0.1}*$ \\
        H$_2$ dissociation pressure & $\log \rm P_{HI}$ & Uniform(-6, 2) & 1.7 & $-1.6^{+2.3}_{-2.7}*$ \\
        Scale factor & scale & TruncLogNormal(0, 0.1, 7.5) & 0.08 & $0.04^{+0.09}_{-0.08}$ \\
        \smallskip \\
         & & \textbf{Derived Parameters} & \\
        \hline
        Carbon/oxygen ratio & C/O & - & 0.01 & $0.1^{+0.4}_{-0.1}* (<0.9)$ \\
        Carbon abundance & [C/H] & - & -0.4 & $0.1^{+0.6}_{-0.5}$ \\
        Oxygen abundance & [O/H] & - & 1.3 & $1.1\pm0.8* (>-0.1)$ \\
        Volatile abundance & [(C+O)/H] & - & 1.1 & $1.0\pm0.7*(>-0.15)$ \\
    \enddata 
    \caption{List of parameters, priors, and results for atmospheric retrievals. The results columns show the values for the 6-PC retrieval. The error bars on the retrieved medians correspond to the 68$\% / 1\sigma$ confidence interval. Parameters with poorly constrained posteriors are indicated with a $*$, with upper/lower 95\% confidence intervals in parentheses if available. In addition to these priors, we required that the atmospheric temperature stay below 6000 K at all pressure levels. The full corner plot is included in Appendix \ref{app:corner}.}
    \label{tab:priors}
\end{deluxetable*}

\begin{figure*}
    \centering
    \includegraphics[width=0.9\linewidth]{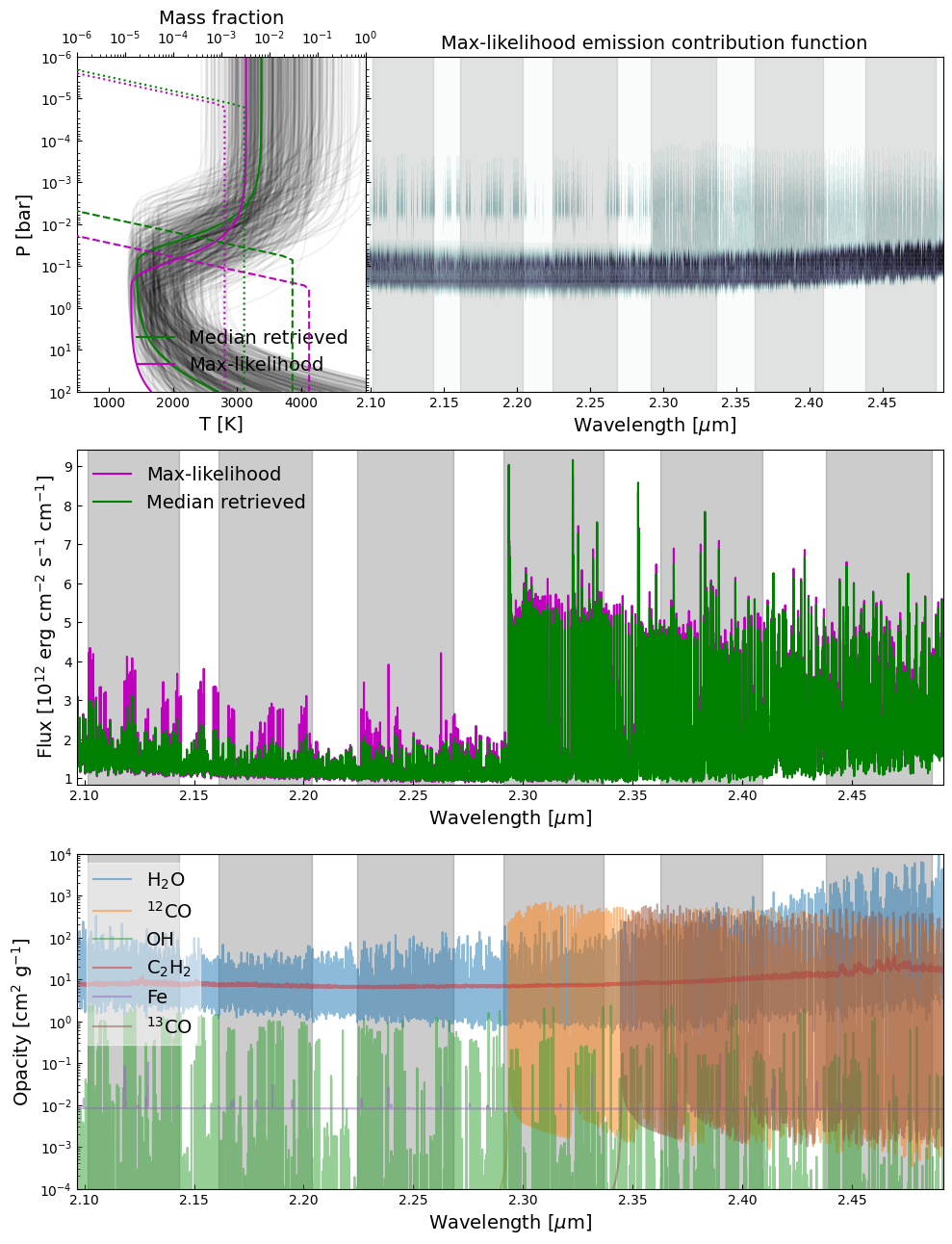}
    \caption{Retrieved $P-T$ profiles (top left), maximum-likelihood emission contribution function (top right), median and maximum-likelihood spectra (middle), and species opacities (bottom). For the $P-T$ profile we show the profile of the median retrieved parameters, the maximum-likelihood profile, and the profiles obtained from 500 draws from the retrieved posterior. The vertical abundance profiles for H$_2$O and CO in each model are shown as dashed and dotted lines (respectively) of the corresponding color. The emission contribution function indicates our observations are sensitive to a wide range of pressures from 1 bar to $\sim10\mu\rm bar$. The maximum-likelihood spectrum is dominated by CO features beyond 2.3 $\mu$m, with weaker OH and H$_2$O features present at bluer wavelengths. The gap in the emission contribution function seen in the first three orders is a result of H$_2$O dissociation at higher pressures, followed by OH formation at lower pressures. We discuss this behavior in Section \ref{ssec:dissoc} Observed NIRSPEC orders are shaded in grey. From the opacity plot, we expect our observations will be dominated by H$_2$O and CO features, with lower sensitivity to OH and C$_2$H$_2$ due to the weaker spectral features of these molecules in the $K$-band.}
    \label{fig:specplot}
\end{figure*}

Results from the 6-component retrieval are presented in Table \ref{tab:priors} and Figure \ref{fig:specplot}. Parameters marked with a $*$ had poorly constrained posteriors, for these parametes 95\% confidence limits are provided if available. $K_\mathrm{p} - \Delta v_\mathrm{sys}$ plots for each species are presented in Figure \ref{fig:kpvsys}. Full corner plots for all four retrievals are included in Appendix \ref{app:corner}.

\subsection{Velocity and winds}\label{ssec:wind}

\begin{figure}
    \centering
    \includegraphics[width=0.95\linewidth]{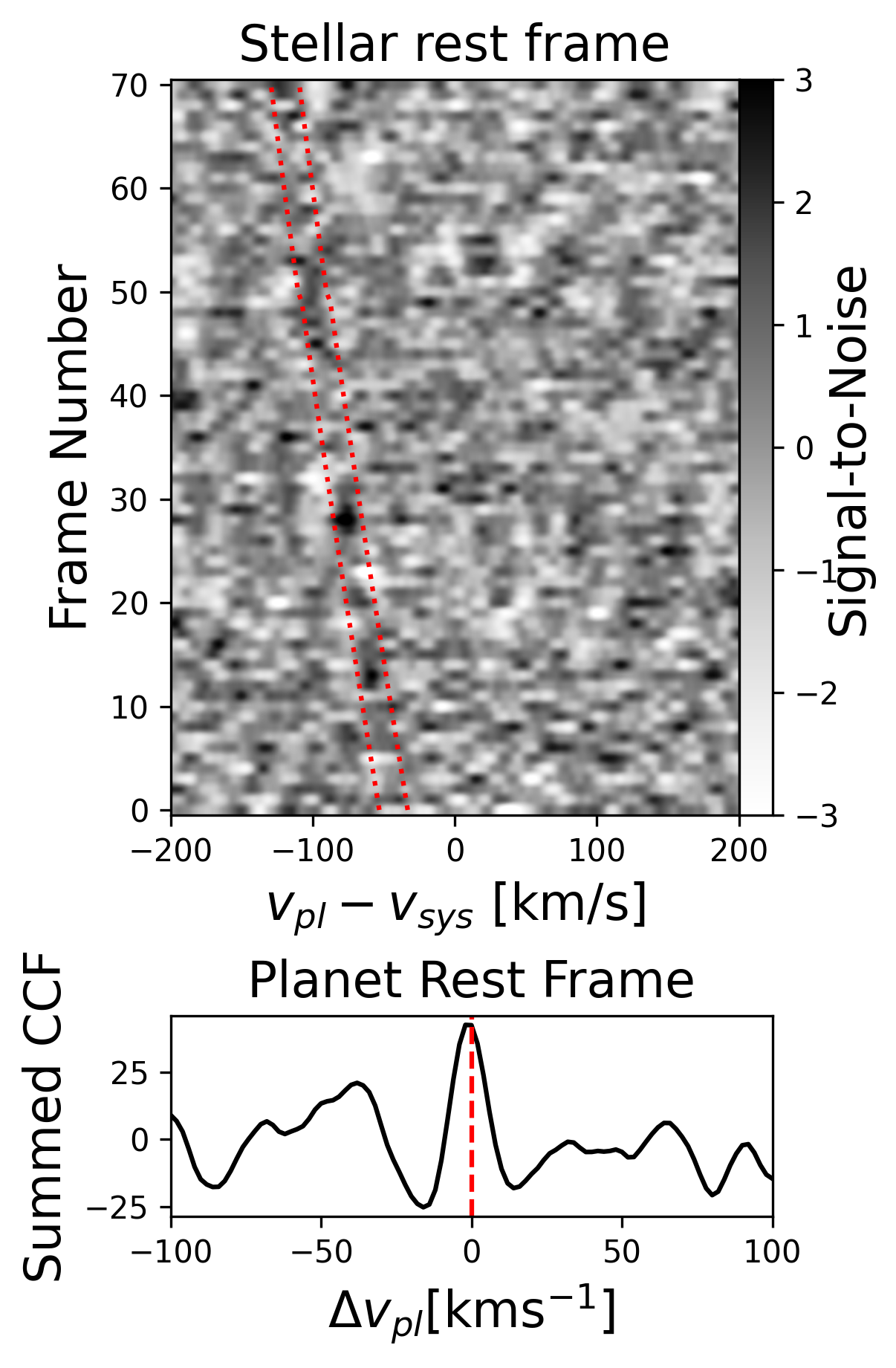}
    \caption{Cross-correlation signal-to-noise as a function of frame number and planet velocity (top) and coadded 1D cross-correlation in the planet rest frame (bottom), using the retrieved maximum-likelihood planet model. The dashed red lines in the top panel indicate $\pm10$ \kms\ from the expected velocity track based on Table \ref{tab:priors}, while the dashed red in the bottom panel indicates the nominal planet rest frame. The planet velocity track is weakly evident in the 2D cross-correlation, and very clear after summing over the time series. No significant velocity offsets are evident, consistent with the retrieved velocity parameters.}
    \label{fig:vtrack}
\end{figure}

The $\Delta K_\mathrm{p}$, $\Delta v_\mathrm{sys}$, and $v_\mathrm{rot}$ parameters are all well-constrained in the retrievals. Both $\Delta K_\mathrm{p}$ and $\Delta v_{\rm sys}$ are consistent with the nominal values in Table \ref{tab:props} within $1\sigma$, and are strongly covariant due to observations being obtained only post-eclipse. Figure \ref{fig:vtrack} shows the cross-correlation as a function of frame number and planet velocity, as well as the cross-correlation function summed in the planet rest frame. The planetary signal is weakly apparent in the 2D plot, but very clear after coadding in the planet rest frame. The planet signal does not appear to be significantly offset from the nominal parameters, consistent with the results of the retrieval analysis.

$v_{\rm rot}$ is well-constrained to $7.5\pm0.7$ \kms, compared with an expected rotational velocity of 2.7 \kms\ assuming tidal locking. This value is physically plausible for a tidally-locked UHJ with a super-rotating equatorial jet \citep[e.g.][]{komacek2022} and is consistent with recent dayside wind measurements from other UHJs \citep{lesjack2025}. This suggests that the LSF fitting procedure described in Section \ref{sec:obs} can provide meaningful constraints even for slowly rotating ($v\sin i < 10$ \kms) planets, rather than being restricted to the young, rapidly-rotating population probed by direct imaging targets. However, we note that we assumed a fixed factor of 1.14 when converting from the spatial to the spectral LSF widths, potentially resulting in a systematic error if this value is incorrect. We also note that the NIRSPEC velocity resolution is itself $\sim9$ \kms. The retrieved value for $v\sin i$ corresponds to a $\sim30\%$ broadening of the intrinsic LSF, assuming the kernel widths add in quadrature. Given the systematic uncertainty associated with the spatial/spectral correction, we caution that this apparent constraint may not correspond to the actual rotational velocity of \keltb. 

\subsection{Pressure-temperature profile}\label{ssec:thermal}

The top row of Figure \ref{fig:specplot} shows the retrieved $P-T$ profile and maximum-likelihood emission contribution function. The emission contribution function indicates our observations are sensitive to emission arising between 50 mbar and 100 $\mu\rm bar$. This is consistent with our limite ability to constrain the intrinsic temperature parameter ($T_{\rm int})$ of the \citet{guillot2010} profile, which mostly impacts the $P-T$ profile at pressures greater than 1  bar. 

The retrieved $\rm T_{eq}$ runs to the lower end of the prior range. Tests with a wider prior found this is remains the case even if the prior is extended to 800 K, but that increases in the multiplicative scale factor applied to the planet model keep the final star/planet flux ratio in the forward model approximately constant. This indicates that the use of a log-normal scale factor prior and tight temperature prior is not sufficient to break the scale factor/temperature degeneracy for small-bandpass high-resolution observations discussed in \citet{finnerty2024}. Resolving this degeneracy appears to require a significantly wider bandpass than our observations cover \citep[such as provided by Gemini/IGRINS, e.g.][]{line2021}, over which the planetary SED shows significant curvature. This leads to a non-linear variation in the planet/star flux ratio $F_p/F_s$ as a function of wavelength which cannot be fit by changes to the scale factor, instead pushing the retrieval to a $P-T$ profile with the correct continuum shape and absolute temperature.

For the presented retrievals, we chose a relatively tight prior bound that leads to a scale factor of approximately 1. We discuss the degeneracies in the absolute abundances in more detail in Section \ref{sec:disc}

As our observations are not flux-calibrated, we are actually fitting for the flux of the planet relative to the host star flux ($F_p/F_s$) using a fixed model of the stellar spectrum. This means that errors in the wavelength-dependent stellar spectrum can potentially bias the retrieved absolute planet temperature and/or scale factor. While we neglected the impacts of stellar oblateness or gravity darkening in our stellar model, the reported $v\sin i = 114$ \kms\ value for KELT-20~A \citep{talens2018} suggests these effects may be significant. This may lead to a bias in the retrieved temperature and/or continuum level in order to correctly match the observed $F_p/F_s$ with an inaccurate $F_s$.

We also note that our observations were taken post-eclipse. Thus, the observed planetary disk is increasingly dominated by the morning limb over the course of our data. This phase coverage may also impact the retrieved temperature and apparent flux of the planet, as the morning/pre-dawn longitudes are expected to be substantially cooler than the dayside average. 

\subsection{Chemical composition}\label{ssec:chem}

\begin{figure*}
    \centering
    \includegraphics[width=0.9\linewidth]{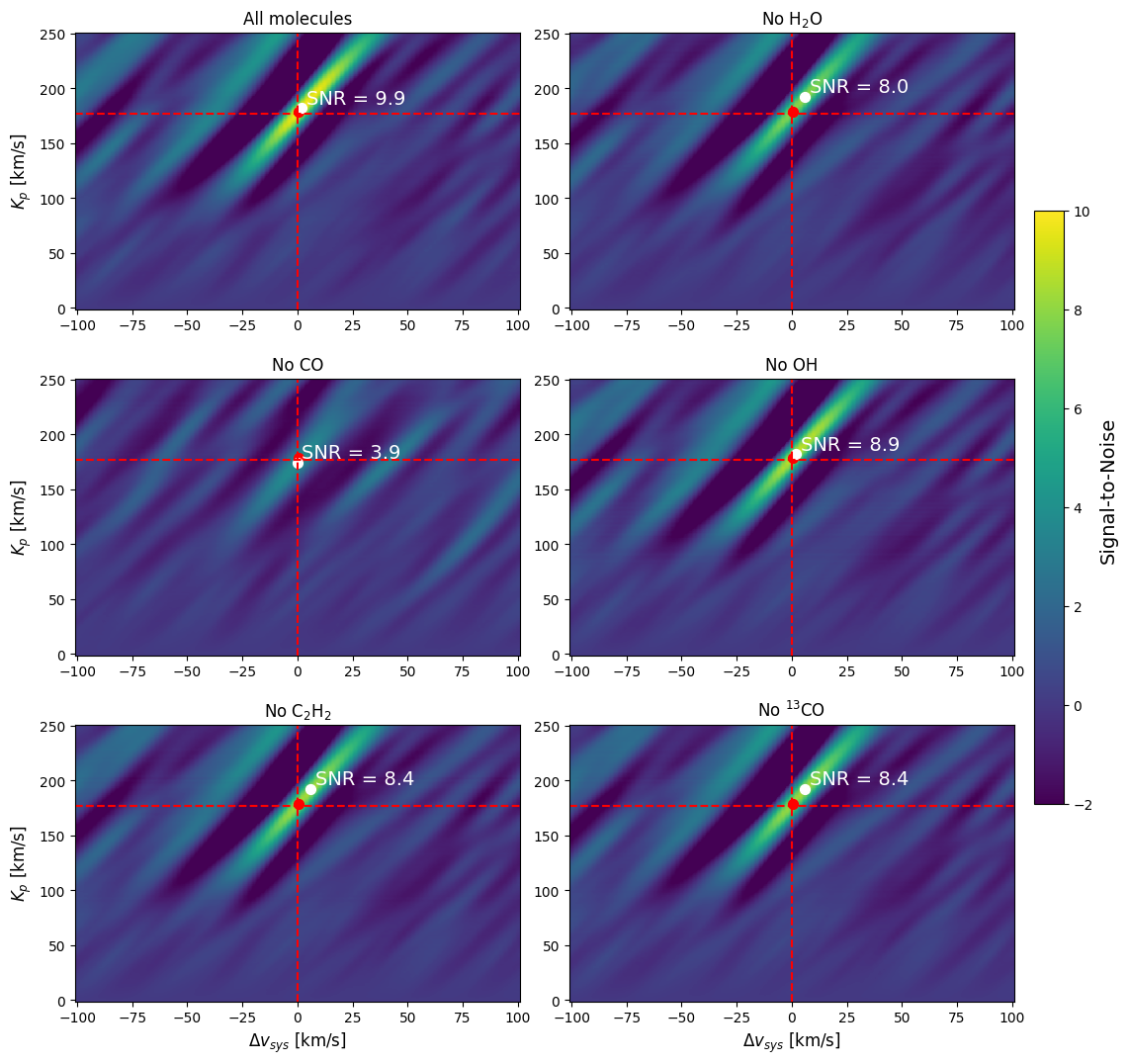}
    \caption{$K_\mathrm{p} - \Delta v_{\rm sys}$ diagrams for the maximum-likelihood model in the top left, with $K_\mathrm{p} - \Delta v_{\rm sys}$ diagrams made with the maximum-likelihood model omitting the specified species in the remaining panels. We use an ad-hoc approach to estimating the detection significance, first subtracting each $K_\mathrm{p}$ column by its median and then dividing by the standard deviation of the $K_\mathrm{p} < -40$ \kms\ region in order to minimize impacts of the detrending process removing the planet signal at small $K_\mathrm{p}$. See \citet{finnerty2024} for a more detailed discussion. Omitting each molecule one-by-one allows us to assess to contribution of that molecule to the overall detection while minimizing the change to the pseudocontinuum that results from omitting a molecule. The total detection is dominated by CO, with a significant contribution from H$_2$O. Including OH, C$_2$H$_2$, and $^{13}$CO has only a marginal impact on the detection strength. These detection strengths are consistent with expectations based on the abundance constraints (or lack thereof) reported in Table \ref{tab:priors}.}
    \label{fig:kpvsys}
\end{figure*}

Our retrievals successfully constrain the CO abundance, and place a lower limit on the H$_2$O abundance. The remaining species in the retrieval are unconstrained, although the OH and Fe posteriors show weak peaks and large C$_2$H$_2$ abundances are weakly disfavored. This is consistent with expectations based on the $K$-band opacities (see Figure \ref{fig:specplot}, bottom panel), which are dominated by CO and H$_2$O features, as well as with previous infrared observations of UHJs \citep[e.g.][]{finnerty2023}. The $K_\mathrm{p} - \Delta v_{\rm sys}$ plots for each molecule using the maximum-likelihood model are shown in Figure \ref{fig:kpvsys}. The contribution of each molecule towards the total detection strength is consistent with the expectation that species which are better-constrained by the retrievals are more significantly detected in the $K_\mathrm{p} - \Delta v_{\rm sys}$ plot.

We also obtain a bounded constraint on the H$_2$O dissociation pressure, $\log P_{H_{2}O} = -1.2^{+0.5}_{-0.7}$ bar. We discuss this constraint and its comparison to chemical equilibrium models in Section \ref{sec:disc}. Key here is that the retrieved H$_2$O dissociation pressure shows a significant degeneracy with that of the CO abundance, with lower dissociation pressure corresponding to higher CO abundances. In previous HRCCS studies, molecular abundances are often correlated, as the loss of continuum information during data processing makes HRCCS retrievals more sensitive to line ratios than absolute line strengths \citep[e.g.][]{finnerty2023, finnerty2024}. In this case, the H$_2$O feature strength appears to be set primarily by the dissociation pressure, rather than the H$_2$O abundance, and so the degeneracy is between the CO abundance and H$_2$O dissociation pressure rather than the abundances themselves.

\subsubsection{Significance of other species}

H$_2$O dissociation is expected to produce significant quantities of OH at pressures where H$_2$O is being dissociated. The retrieved posteriors do not provide bounded constraints on the OH abundance or the OH dissociation/formation pressures, leading to an unconstrained vertical mixing profile shown in Figure \ref{fig:chemplot}. Figure \ref{fig:specplot} demonstrates the comparatively low OH opacity in the $K$-band relative to other species, which is likely the reason for the poor constraints. The insignificance of OH to the detection is further demonstrated by Figure \ref{fig:kpvsys}, which shows that omitting OH from the maximum-likelihood model has only a marginal impact on the detection strength in the $K_p - \Delta v_{sys}$ space. 

The retrieved posterior weakly disfavors high C$_2$H$_2$ abundance, consistent with the weakness of high-temperature C$_2$H$_2$ spectral features in the $K$-band. C$_2$H$_2$ was included in the retrievals because it can be present in significant, potentially detectable quantities for C/O ratios greater than 1, but our retrievals strongly disfavor this scenario (see Section \ref{sec:disc}). Similarly with OH, omitting C$_2$H$_2$ entirely from the maximum-likelihood planet model has only a minimal impact on the overall detection strength. 

\subsection{CO Isotopologue Ratio}
Finally, the retrieved posterior places an upper limit on the CO isotopologue ratio, $\rm \log ^{13}CO / \log ^{12}CO < -2.0$ at 95\% confidence. This is only slightly below the solar-system value of 89, and stands in contrast with previous results from KPIC observations of hot Jupiters suggesting significant $^{13}$CO enrichment \citep{finnerty2023,finnerty2024}. 

\subsection{Number of principal components}\label{ssec:ncomps}

\begin{figure*}
    \centering
    \includegraphics[width=0.9\linewidth]{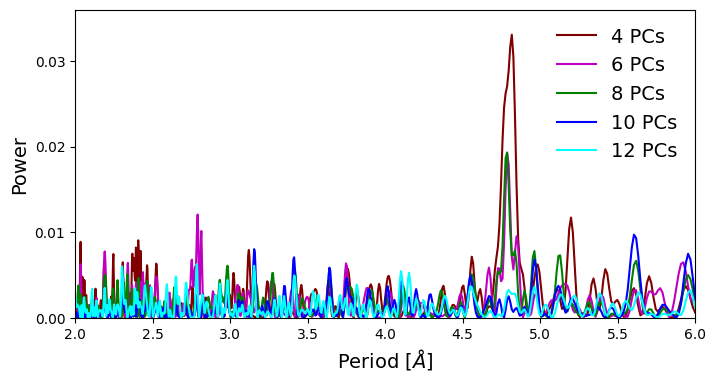}
    \caption{Lomb-Scargle periodograms of the median spectrum for order 31 after detrending. A significant peak at $\sim4.8\AA$ is consistent with incomplete removal of the KPIC fringing effect described in \citet{Finnerty2022}. The power present in this peak is significantly reduced by projecting out at least six principal components. This suggests sufficient removal of fringing/telluric effects in this case requires omitting at least six components, and that the odd posterior behavior seen in the four component retrievals may be a result of undercleaning. Removing 10 or more components is required to entirely eliminate this fringing signal, but these retrievals also give significantly broader posteriors, suggesting that the planet signal is being impacted by omitting so many components.}
    \label{fig:pcplot}
\end{figure*}

We ran retrievals omitting 4, 6, 8 and 12 principal components in order to assess the impact of the number of dropped principal components on the retrieved posteriors. The full corner plots for each retrieval are included in Appendix \ref{app:corner}. Based on inspection of the corner plots and fringing periodograms, we adopt the 6 component case as our fiducial retrieval. The 4 component retrieval provides an unexpectedly tight constraint on the Fe abundance, and strongly prefers low values of $\rm T_{eq}$, which are counteracted with a larger scale factor. The 8 and 12 component cases yield marginalized posteriors which are mostly compatible with the 6 component case, but significantly wider. In particular, the 8 and 12 component cases prefer higher temperatures, do not constrain the the H$_2$O abundance or dissociation pressure, and give worse constraints on $K_p$ and $v_{sys}$. This suggests that in the 8 and 12 component cases the PCA is distorting/removing portions of the underlying planet spectrum, leading to worse constraints.



To better assess the impact of the PCA, we visually inspected the post-PCA spectral time series. This revealed clear evidence of residual fringing which has not been removed by omitting only 4 principal components, particularly at the beginning and end of the observation. To confirm this, we computed the Lomb-Scargle periodogram of the cleaned time series for order 31 after omitting 4, 6, 8, 10, and 12 principal components, shown in Figure \ref{fig:pcplot}. A clear peak is seen at a period of 4.8$\rm \AA$, likely corresponding to the KPIC dichroic fringing effect described in \citep{Finnerty2022}. This peak is substantially stronger when omitting only 4 components than in the 6 and 8 component cases, suggesting that the 4 component retrieval may be impacted by unremoved fringing. The fringe feature disappears from the periodogram when omitting 10 or more components, but comparing the 6 and 8 component retrievals suggests that the PCA is already beginning to significantly degrade the planet spectrum in the 8 component case. As a result, we choose the 6 component case as our fiducial retrieval for Table \ref{tab:priors} and subsequent discussion.

\section{Discussion}\label{sec:disc}

\subsection{Atmospheric circulation}

Our median retrieved values for $\Delta K_\mathrm{p}$ and $\Delta v_{\rm sys}$ are compatible with the orbital ephemeris and assumed velocity parameters listed in Table \ref{tab:props}. This suggests that the detected planetary emission is not dominated by one limb of the planet, which would lead to a net blueshift if the morning limb dominated or a redshift if the evening limb were dominating the total flux as a result of the planet's rotation.

While the retrieved velocity parameters are consistent with the \citet{exofop3} ephemeris, using the values from \citet{talens2018} instead results in a $\sim4$ \kms\ blueshift. Alternatively, using period/transit time values from \citet{lund2017, patel2022, ivshina2022}, or \citet{kokori23} change the nominal planet velocities by $\sim1-3$ \kms. In some cases, propagating the reported uncertainties on the period and transit time is insufficient to explain this difference. 

This discrepancy speaks to the importance of accurate knowledge of the planetary ephemeris for interpreting  offsets in the $K_p - \Delta v_{sys}$ diagram as a result of atmospheric circulation pattern.  Inaccuracies as small as 10 minutes in the mid-transit time can lead to 10 \kms\ offsets in the planet velocity in some systems \citep{pai2022}, comparable to the fastest winds expected on UHJs. While transmission observations can use ingress/egress to very precisely constrain the ephemeris at the observed epoch, emission observations must rely on a preexisting ephemeris with its associated uncertainties. As a result, we suggest any attempts to interpret velocity offsets in emission observations proceed cautiously. 

In the case of \keltb, additional difficulties in interpreting velocity offsets arise from uncertainty in the systemic velocity of the rapidly-rotating A-type primary.  \citet{talens2018} reports a value of $-21.3\pm0.4$\kms, while \citet{lund2017} gives a value of $-23.3\pm0.3$\kms. \citet{johnson2023} report a value of $-26.0$\kms, and \citet{stangret2024} report a value of $-24.6$ \kms. This uncertainty is comparable to the expected wind velocities for UHJs, precluding robust measurement of wind speeds even from transmission observations \citep{stangret2024}. 

Rotational broadening of planetary lines is not expected to be sensitive to ephemeris or stellar velocity errors, providing an alternative probe of global circulation patterns for emission observations when the instrumental LSF is well-characterized, as is the case for KPIC. If the observed spectrum is dominated by the morning or evening limb of the planet, the rotation will lead to a net redshift or blueshift of the planetary atmosphere, respectively. Similarly, a strongly longitudinally-dependent wind pattern may lead to a net red/blueshift. In contrast, a symmetric eastward flowing equatorial jet would not lead to a net shift in the observed planet spectrum, but may instead appear to increase the rotational velocity of the planet beyond the expected value based on tidal locking.

Assuming a tidal locking and the parameters listed in Table \ref{tab:props}, the expected rotational velocity of \keltb\ is 2.7 \kms. The retrieved value of $7.5\pm0.7$ \kms\ could be explained by a 5 \kms\ eastward-flowing equatorial jet dominating the observed emission and appearing to increase the rotational velocity of the planet. This is consistent with wind speeds previously reported for \keltb\ \citep{casasayas2019, nugroho2020, stangret2020, petz2024}, and with model predictions for UHJs \citep[e.g.][]{komacek2022, beltz2022}. 

The retrieved rotational broadening is comparable to the instrumental resolution of NIRSPEC, increasing the line widths by only $\sim30\%$ compared to the intrinsic instrumental LSF. While the large number of planet lines and the well-characterized LSF is reason to believe we can measure line broadening to the $\sim10\%$ level, systematic uncertainties in the spatial/spectral profile conversion \citep{Finnerty2022} suggest the retrieved $v\sin i$ should be taken with caution. Unambiguous measurement of $v\sin i$ may require higher spectral resolution, which will be available with Keck/HISPEC \citep{hispec}.  

\subsection{Molecular Dissociation}\label{ssec:dissoc}

\begin{figure*}
    \centering
    \includegraphics[width=0.9\linewidth]{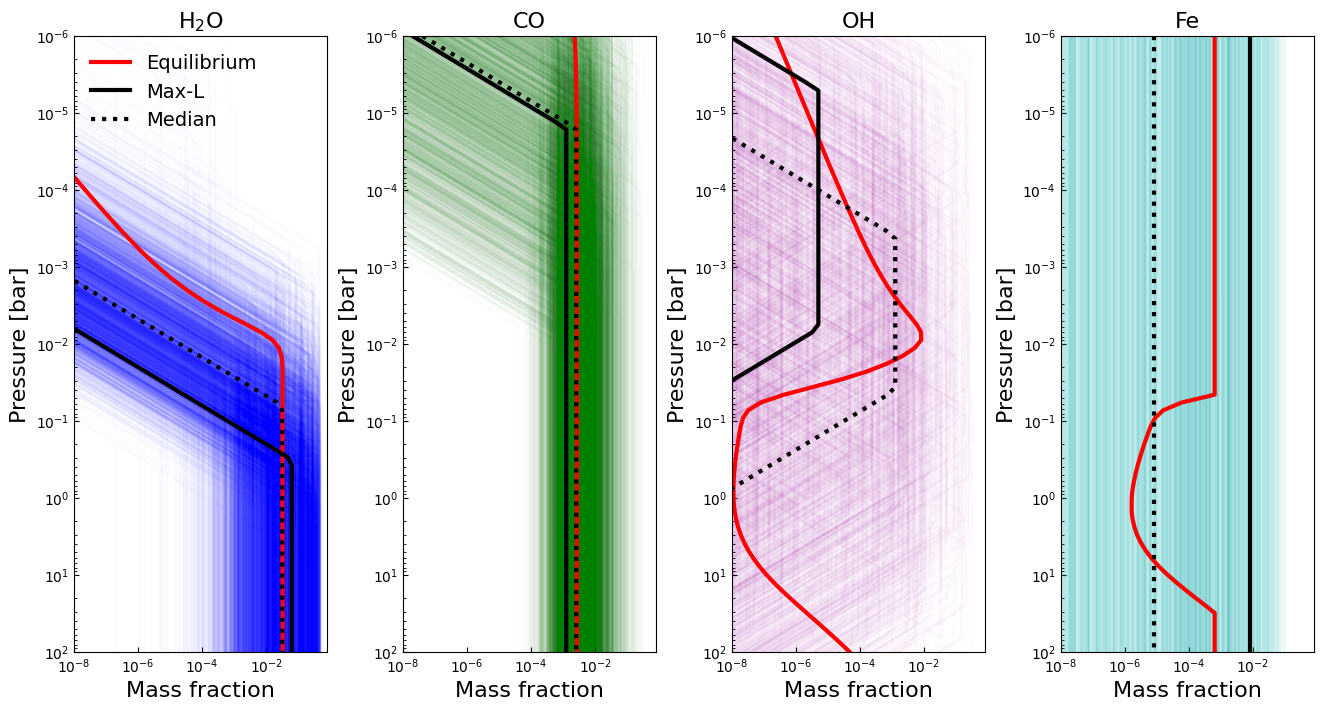}
    \caption{Retrieved abundance profiles compared with compared with \texttt{easychem} equilibrium predictions. The colored profiles are 4000 draws from the posterior, while the maximum-likelihood and median profiles are indicated in solid and dotted black, respectively. The equilibrium profiles are shown in solid red. For the equilibrium model, we used the $P-T$ profile corresponding to the median retrieved values. The median abundance profiles for H$_2$O and CO are roughly consistent with the equilibrium model for C/O = 0.05 and [Fe/H] = -0.3 (shown in the plot), although the retrieval prefers somewhat deeper dissociation of H$_2$O. These values are significantly lower than those given in Table \ref{tab:priors} due to the inclusion of refractory species in the equilibrium model. The OH and Fe profiles are poorly constrained, consistent with the overall insignificance of the detection.}
    \label{fig:chemplot}
\end{figure*}

We plot the median and maximum-likelihood vertical abundance profiles compared with chemical equilibrium predictions from \texttt{easyCHEM} \citep{molliere2017, lei2024} in Figure \ref{fig:chemplot}. For the equilibrium prediction, we use the median $P-T$ profile parameters. The C/O ratio and metallicity of the equilibrium model were then adjusted by eye to roughly match the median abundance profiles, which we discuss further in the next subsection. We find a reasonable agreement between the retrieved abundance profiles and the chemical equilibrium predictions for C/O $\sim0.05$ and [Fe/H] = $\sim-0.3$. We discus the discrepancy between this value and our retrieved [(C+O)/H] in the following subsection.  

The H$_2$O abundance profile from the chemical equilibrium model dissociates moderately higher in the atmosphere than the abundance profile preferred by the retrieval, but is within the range of the posterior. This concurrence suggests that chemical equilibrium is a good assumption when considering only CO and H$_2$O for \keltb\, and assuming equilibrium may help break the previously-discussed temperature/scale factor degeneracy while reducing the number of free parameters in the retrieval. At the same time, this agreement demonstrates that free retrievals can account for vertically varying abundances with only 1-2 additional free parameters per molecule, while also retaining a flexibility for possible photochemistry.

Consistent with the non-detection of OH in Figure \ref{fig:kpvsys}, the OH abundance profile is unconstrained. The poor OH constraint is due to the relatively low opacity of OH in the $K$-band and is consistent with previous KPIC $K$-band observations of WASP-33~b \citep{finnerty2023}. OH has been detected in $H$-band observations of UHJs \citep{nugroho2021, landman2021, brogi2023, weinermansfield2024} due to the higher line opacity in that band. Combining $H$-band observations which constrain the abundance profiles for OH and H$_2$O with $K$-band observations constraining the ratio of H$_2$O features to CO features may provide much better constraints on the bulk C/O ratio in the presence of H$_2$O dissociation than is possible from $K$-band observations alone. Such observations will be possible with Keck/HISPEC, which will provide high-resolution fiber-fed spectroscopy covering the full 1-2.5 $\mu$m range in a single shot \citep{hispec}. 

\subsection{Interactions between P-T and mixing profiles}

While experimenting with different parameterizations of the $P-T$ and dissociation profiles, we encountered complicated degeneracies in the resulting posteriors and unusual behavior in the retrieved $P-T$ profiles. These complications all arise from the presence of two terms impacting the radiative transfer at each level of the model atmosphere, rather than one. Beginning from Kerckoff's Law of radiative transfer, we have:

\begin{equation}
    \frac{d I_\nu}{ds} = -n(s)\sigma_\nu I_\nu + n(s)\sigma_\nu B_\nu(T)
\end{equation}

Where $I_\nu$ is the specific intensity, $n(s)$ is the number density of an absorbing species at altitude $s$, $\sigma_\nu$ is the absorption cross section of the species, and $B_\nu(T)$ is the Planck function evaluated at the temperature of altitude $s$. In the case of abundances fixed with altitude, the change in intensity is set only by the change in $B_\nu(T)$ with height, resulting in an emission contribution function which tracks the $P-T$ profile. However, in the case of vertically-varying abundances, $I_\nu$ can change even with an isothermal $P-T$ profile as a result of changes in $n(s)$. Depending on the specifics on the $P-T$ and mixing profiles, this can lead to emission contributions functions with multiple layers and/or gaps which may or may not correspond to regions of curvature in the $P-T$ profile.

An example of this can be seen in the upper right panel of Figure \ref{fig:specplot} showing the retrieved maximum-likelihood emission contribution function. In the bluest three orders, the emission contribution function drops off rapidly when H$_2$O dissociation begins, only to increase again at lower pressures as the model includes increasing quantities of OH. This increase occurs even as the curvature of the $P-T$ profile rapidly decreases in this pressure range. 

A more straightforward effect is introduced by fixing the intrinsic temperature parameter $\rm T_{int}$, which leads to a multimodality between the H$_2$O abundance and dissociation pressure. Retrievals with $\rm T_{int} = 100\ K$ returned posteriors with either a low H$_2$O abundance and dissociation at $\rm \sim1\ mbar$ or a high H$_2$O abundance and dissociation at $\rm \sim 100\ mbar$. The planet spectra in each mode had similar H$_2$O feature strength, leading our free retrieval approach to not significantly prefer one mode over the other. Including $\rm T_{int}$ as a free parameter with a lower prior bound high enough to prevent an isothermal lower atmosphere appears to break this degeneracy in favor of the high-abundance, deep-dissociation mode. Future HRCCS analyses including molecular dissociation should take take care to assess the precise impact of their $P-T$ parametrization on the final posterior. 

As a consistency check, we also ran retrievals using a physics-free spline-based $P-T$ parametrization as described by \citet{pelletier2021}. The resulting $P-T$ profile reached a minimum of $\sim 1000$ K between 1 and 0.1 bar, but the abundance/dissociation parameters were consistent with the presented values. This suggests that changes in the $P-T$ parametrization/limits are not strongly impacting the retrieved abundances. Using the spline-based approach significantly increases the number of free parameters in the fit, leading to longer runtimes. We also find the spline profile is prone to non-physical oscillations in the upper/lower atmosphere where the emission contribution function is negligible, even for aggressive values of the smoothing prior. We therefore prefer the \citet{guillot2010} approach for this work, but note that analyses covering a wider bandpass will be more favorable to spline-based approaches.   
\subsection{C/O and metallicity}

\begin{figure}
    \centering
    \includegraphics[width=0.95\linewidth]{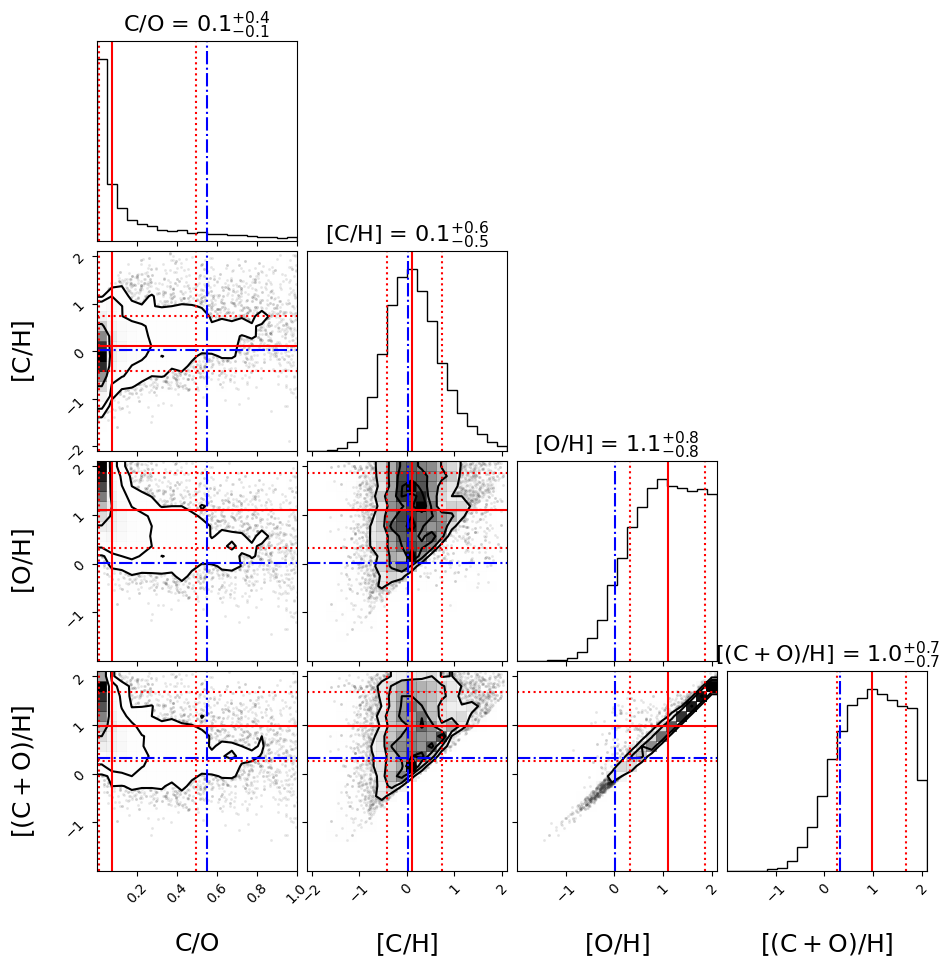}
    \caption{Corner plot of C/O, [C/H], [O/H], and [(C+O)/H] from the retrieved posterior. The medians are in solid red, with dashed red indicating the $\pm34$\% confidence intervals. The solar values from \citet{solarabunds} are shown in dash-dot blue. The carbon abundance is close to the solar abundance, while [O/H] prefers solar to super solar values ($\rm [O/H]> -0.1$ at 95\% confidence). This leads to a very low C/O ratio and a lower limit on the total volatile abundance $\rm [(C+O)/H ] >-0.15$ at 95\% confidence}
    \label{fig:coch}
\end{figure}

The corner plot for the retrieved C/O, [C/H], [O/H], and [(C+O)/H] values are shown in Figure \ref{fig:coch}, and the median values and confidence intervals are listed in Table \ref{tab:priors}. These posteriors were constructed from the retrieved abundances of CO, H$_2$O, and C$_2$H$_2$ in order to estimate the deep atmosphere C/O ratio below any molecular dissociation, and do not account for contributions from any other species. OH is specifically omitted because it is expected to be present only in the upper atmosphere above H$_2$O dissociation, and including both H$_2$O and OH would therefore double count that oxygen when determining the bulk abundances.

As \petit\ uses mass fractions for molecular abundances, we have thus far reported results as mass fractions. However, the C/O, O/H, and C/H values are generally reported as number fractions/volume mixing ratios. The mass fraction can be converted to the volume mixing ratio as follows:

\begin{equation}
    n_X = \frac{\mu}{\mu_X}M_X
\end{equation}

Where $n_X$ is the volume mixing ratio of species X, $M_X$ is the mass fraction of species X, $\mu_X$ is the molar mass of species X, and $\mu$ is the mean molecular weight. The dependence on the mean molecular weight cancels when computing abundance ratios. For logarithmic mass fractions and a fixed mean molecular weight of 2.3 g/mol, $\log \rm H_2O_{VMR} = \log \rm H_2O_{MF} - 0.9$, and $\log \rm CO_{VMR} = \log \rm CO_{MF} - 1.1$

The median and $\pm34$\% confidence intervals are $\rm C/O = 0.1^{+0.4}_{-0.1}$ ($\rm C/O < 0.9$ at 95\% confidence), $\log \rm [C/H] = 0.1^{+0.5}_{-0.4}$, $\log \rm [O/H] = 1.1\pm0.8$ ($\rm [O/H] > -0.1$ at 95\% confidence), and $\rm [(C+O)/H] = 1.0\pm0.7$ ($\rm [(C+O)/H] > -0.15$ at 95\% confidence). As we retrieve only a lower limit for the H$_2$O abundance, the oxygen abundance is similarly a lower limit, as is the total volatile abundance. Solar values for [O/H] and [(C+O)/H] from \citet{solarabunds} are disfavored at $\sim1\sigma$, with the maximum-likelihood parameters corresponding to a $\sim10\times$ solar enrichment in oxygen. The C/O ratio posterior has a sharp cutoff at $\rm C/O = 1$ due to the preference for little/no C$_2$H$_2$ in the atmosphere, which leaves CO as the only carbon species in the atmosphere. 

The C/O and abundance values we report are based only on the molecular species included in our retrievals, and therefore may not be an accurate estimate for the atmosphere as a whole. To address this, we compare the retrieved abundances to equilibrium chemistry models from \texttt{easyCHEM} \citep{molliere2017, lei2024}, shown in Figure \ref{fig:chemplot}. These models scale the atomic abundances of all species except oxygen according to a single metallicity parameter, while the oxygen abundance is determined by a separate parameter setting the C/O ratio. Figure \ref{fig:chemplot} compares the retrieved mixing profiles to an \texttt{easyCHEM} model with C/O = 0.05, [Fe/H] = --0.3, which agrees well with the lower-atmosphere abundances.  While this C/O ratio is comparable to those in our retrievals, the metallicity is substantially lower than the $10\times$ solar volatile abundance we retrieve. The retrieved volatile enrichment is due entirely to the high retrieved oxygen abundance, which in the equilibrium models is determined by the C/O ratio once the metalicity parameter is tuned to match the carbon abundance. As a result, the equilibrium [Fe/H] parameter is not necessarily measuring the bulk metallicity in the event of a refractory/volatile ratio differing from solar. 

At first glance, the low-C/O ratio composition appears to be inconsistent with the previously reported solar or slightly supersolar C/O and atmospheric metallicity for \keltb\ from \citet{kasper2023}, who combined Spitzer, HST, and MAROON-X observations to perform a joint high/low spectral resolution retrieval. However, we note that there is a strong degeneracy in our retrievals between the CO abundance and the H$_2$O dissociation pressure, and that our retrievals prefer H$_2$O dissociation occurring deeper in the atmosphere than expected based on chemical equilibrium. Dissociation closer to the equilibrium expectation corresponds to an enhancement in the CO abundance, leading to a solar or slightly super-solar metallicity and increasing the C/O ratio to approximately the solar value, which is at the edge of the retrieved $1\sigma$ confidence interval for the C/O ratio.  

Observations with broader wavelength coverage could reduce uncertainty in the atmospheric metallicity by observing multiple bands from the same species, which would break the degeneracy between H$_2$O and CO abundances seen when retrieving from $K$-band data alone. Broader wavelength coverage would also allow the detection of additional species, including refractory elements, which would constrain the atmospheric refractory/volatile ratio, which is potentially an important tracer of the planet formation process \citep{lothringer2021, chachan2023}, as well as provide a non-volatile probe of the atmospheric metallicity. Refractory/volatile ratio constraints have recently been demonstrated using IGRINS $H$+$K$-band data \citep{smith2024} and also using a combination of $K$-band and optical observations \citep{pelletier2025}. Future joint retrievals combining KPIC and optical observations will provide similar constraints. Such a joint retrieval is outside the scope of the present work. 

Previous studies of UHJs have found that free retrievals with fixed abundances prefer high-metallicity, high-C/O ratio atmospheres, but that using equilibrium models to account for molecular dissociation results in less metal-rich atmospheres with a C/O ratio closer to solar/stellar values \citep[e.g.][]{brogi2023, ramkumar2023}. This bias in free retrievals occurs because when ignoring dissociation, the only way to reduce the strength of H$_2$O features relative to CO is to reduce the H$_2$O abundance and/or increase the CO abundance, increasing the C/O ratio. In this case, we may be seeing the reverse phenomenon, where dissociation below the chemical equilibrium level is driving a higher H$_2$O abundance, lower CO abundance, and low C/O ratio.

Enforcing chemical equilibrium accounts for dissociation, but the equilibrium assumption for hot gas giants is in tension with results from \textit{JWST} transmission spectroscopy preferring significant spectral features from photochemically produced species \citep{tsai2023, dyrek2024}. Our retrieval of the H$_2$O dissociation pressure for \keltb\ indicates that free retrievals can account for dissociation without making an equilibrium assumption, providing an approach that can avoid biases from fixed vertical abundances while maintaining the flexibility to explore novel photochemical processes.

\subsection{Formation history}

While many ultra-hot Jupiters have significant misalignments between the planet's orbital plane and the spin axis of the host star \citep[e.g.][]{collier2010, gaudi2017, talens2018m1, cabot2021}, \keltb\ is extremely well-aligned with its host star's rotation \citep{talens2018, singh2024}. Furthermore, KELT-20A lies above the Kraft break \citep{kraft1967}, in a regime where close-in giant planets are preferentially misaligned \citep{Winn2010} and tidal realignment is likely insignificant over the lifetime of the system \citep{albrecht2012}.  This suggests \keltb\ could possibly have formed without high-eccentricity migration, which is necessary to produce the extremely inclined/retrograde orbits seen in other UHJs \citep{naoz2011}. Differences in the formation history of \keltb\ compared to other UHJs could lead to an anomalous composition compared to the population as a whole. 

The low-C/O, solar-to-supersolar metallicity preferred by our retrievals is consistent with general predictions from core-accretion models that these parameters are inversely correlated in exoplanet atmospheres \citep{espinoza2017, madhusudhan2017, cridland2019}, and specifically with models for formation beyond the CO snow line followed by Type II migration \citep{khorshid2022}. However, our retrievals are ambiguous as to the bulk metallicity of \keltb, as we do not constrain the abundances of any refractory species. Our retrieved composition could be explained by substantial oxygen enrichment with an otherwise low metallicity, or by a carbon-poor but otherwise metal-enriched atmosphere. Further ambiguity arises from the deeper retrieved H$_2$O dissociation pressure compared with chemical equilibrium models, which may lead to an over-estimation of the H$_2$O abundance and corresponding under-estimation of the C/O ratio. Solar or moderately super-solar atmospheric metallicity is consistent with equilibrium retrievals of other UHJs, which have preferred solar C/O ratios for both aligned \citep{brogi2023} and misaligned \citep{ramkumar2023} systems. 

In the absence of isotopologue constraints, measurements from other portions of the optical/infrared spectrum to constrain the refractory/volatile ratio could distinguish between different formation scenarios \citep{lothringer2021, chachan2023}. In the case of \keltb, the extensive existing optical data \citep[e.g.][]{johnson2023, petz2024} may allow joint high-resolution optical/infrared retrievals and constraints on the refractory/volatile ratio without the need for additional observations.

\section{Summary and Conclusions}\label{sec:conc}

We performed atmospheric retrievals on high-resolution post-eclipse $K$-band observations of the ultra-hot Jupiter \keltb/MASCARA-2b. Using a free-retrieval framework including molecular dissociation, we constrain the abundance of CO and place a lower limit on the abundance of H$_2$O, finding the atmosphere of \keltb\ is consistent with an oxygen-rich ($\rm C/O = 0.1^{+0.4}_{-0.1},\ C/O < 0.9$ at 95\% confidence) composition. The retrieved H$_2$O dissociation pressure and molecular abundances are roughly consistent with predictions from equilibrium chemistry models having C/O $\sim$ 0.05 and metallicity $0.5\times$ solar, but the lack of refractory species detections limits our ability to constrain the bulk metallicity. We also constrain the rotational broadening of \keltb\ to $7.5\pm0.7$ \kms. This is $\sim5$ \kms\ larger than the value expected from tidally locked rotation, and is consistent with a supersonic day-to-night wind. Such a wind is consistent with results from other UHJs and with predictions from GCMs, but the relatively low spectral resolution of KPIC results in a significant systematic uncertainty in the retrieved rotational velocity. 

\begin{acknowledgments}
The thank the anonymous referee whose thorough and insightful comments greatly improved this work. L. F. is a member of UAW local 4811. L.F. acknowledges the support of the W.M. Keck Foundation, which also supports development of the KPIC facility data reduction pipeline. The contributed Hoffman2 computing node used for this work was supported by the Heising-Simons Foundation grant \#2020-1821. Funding for KPIC has been provided by the California Institute of Technology, the Jet Propulsion Laboratory, the Heising-Simons Foundation (grants \#2015-129, \#2017-318, \#2019-1312, \#2023-4597, \#2023-4598), the Simons Foundation (through the Caltech Center for Comparative Planetary Evolution), and the NSF under grant AST-1611623. D.E. acknowledges support from the NASA Future Investigators in NASA Earth and Space Science and Technology (FINESST) fellowship under award \#80NSSC19K1423, as well as support from the Keck Visiting Scholars Program (KVSP) to install the Phase II upgrades for KPIC. J.X. acknowledges support from the NASA Future Investigators in NASA Earth and Space Science and Technology (FINESST) award \#80NSSC23K1434.

This work used computational and storage services associated with the Hoffman2 Shared Cluster provided by UCLA Institute for Digital Research and Education’s Research Technology Group. L.F. thanks Briley Lewis for her helpful guide to using Hoffman2, and Paul Molli\`ere for his assistance in adding additional opacities to petitRADTRANS. 

The data presented herein were obtained at the W. M. Keck Observatory, which is operated as a scientific partnership among the California Institute of Technology, the University of California and the National Aeronautics and Space Administration. The Observatory was made possible by the generous financial support of the W. M. Keck Foundation. W. M. Keck Observatory access was supported by Northwestern University and the Center for Interdisciplinary Exploration and Research in Astrophysics (CIERA). The authors wish to recognize and acknowledge the very significant cultural role and reverence that the summit of Mauna Kea has always had within the indigenous Hawaiian community.  We are most fortunate to have the opportunity to conduct observations from this mountain. 

This research has made use of the NASA Exoplanet Archive, which is operated by the California Institute of Technology, under contract with the National Aeronautics and Space Administration under the Exoplanet Exploration Program.

\end{acknowledgments}

%

\vspace{5mm}
\facilities{Keck:II(NIRSPEC/KPIC)}


\software{astropy \citep{astropy:2013, astropy:2018},  
          \dynesty\ \citep{speagle2020},
          \texttt{corner} \citep{corner},
          \petit\ \citep{prt:2019, prt:2020},
          \texttt{numpy} \citep{numpy},
          \texttt{scipy} \citep{scipy2020},
          \texttt{matplotlib} \citep{matplotlib}}


\appendix
\section{Corner Plots}\label{app:corner}
\begin{figure}
    \centering
    \includegraphics[width=1.0\linewidth]{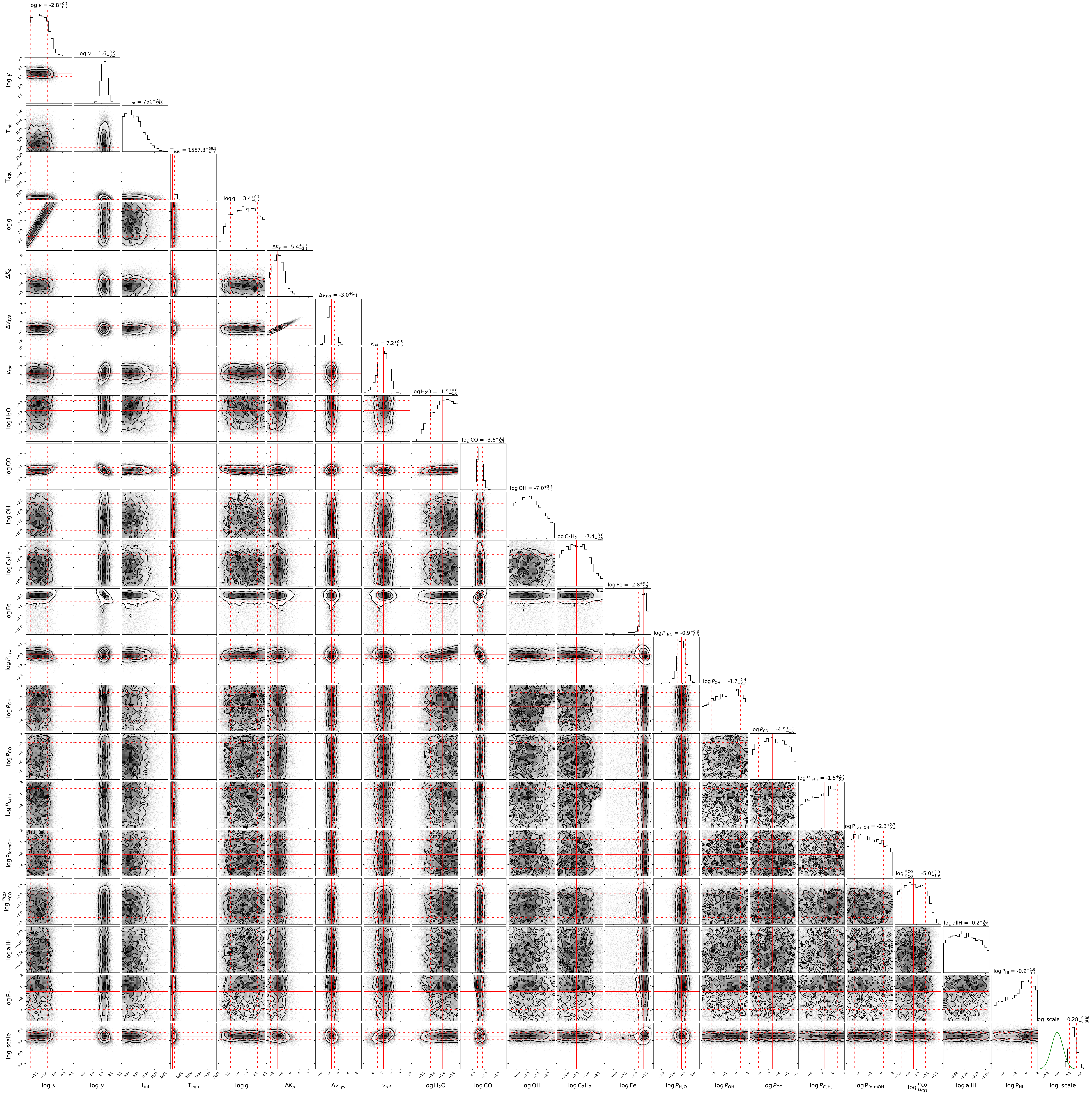}
    \caption{Full corner plot for the retrieval omitting 4 principal components. Red solid lines indicate the medians, while red dashed lines indicate the bounds of the marginalized 68\% confidence interval. We discuss these results in Section \ref{sec:res}.}
    \label{fig:corner4}
\end{figure}

\begin{figure}
    \centering
    \includegraphics[width=1.0\linewidth]{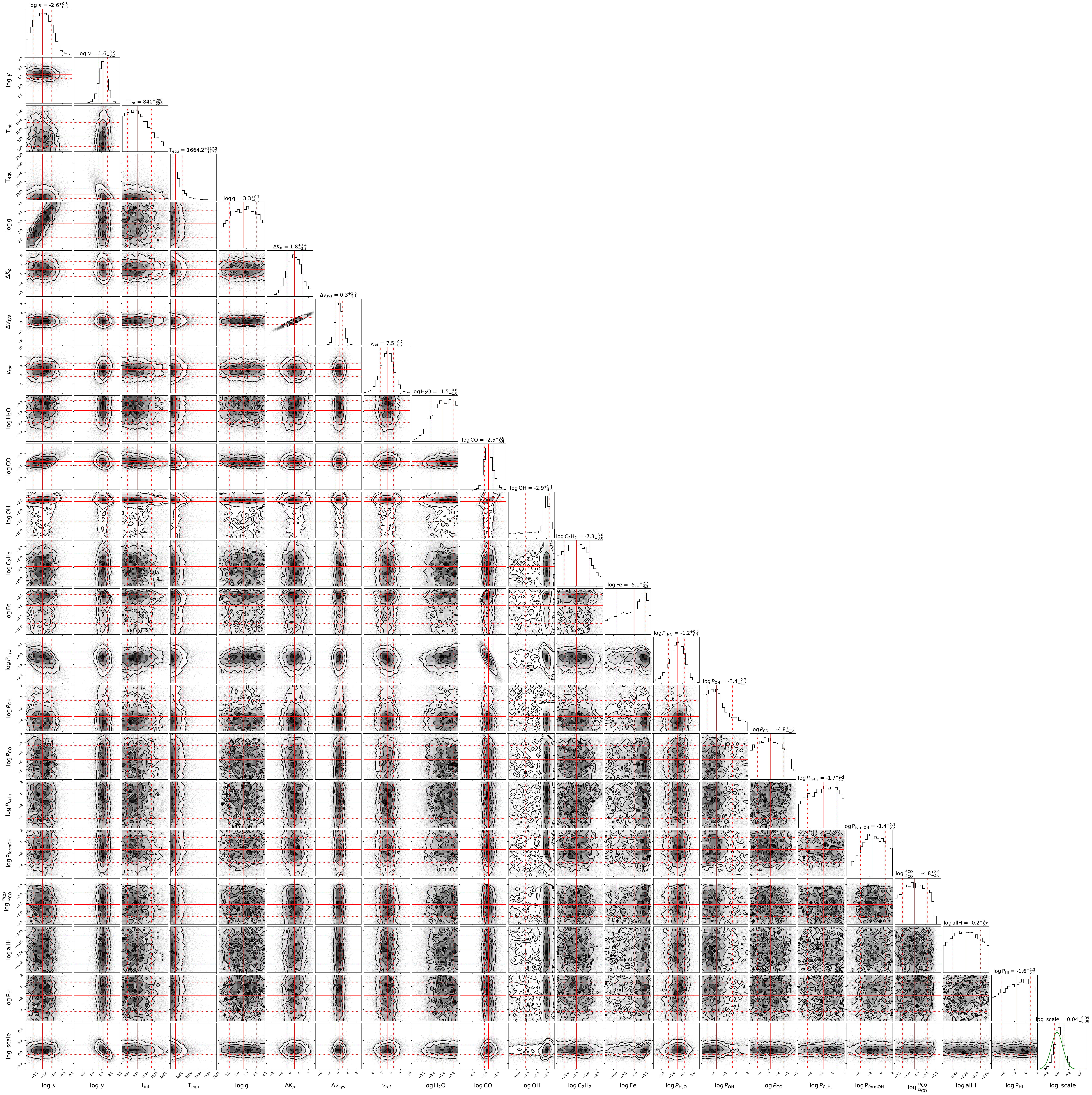}
    \caption{Full corner plot for the retrieval omitting 6 principal components. Red solid lines indicate the medians, while red dashed lines indicate the bounds of the marginalized 68\% confidence interval. We discuss these results in Section \ref{sec:res}.}
    \label{fig:corner6}
\end{figure}

\begin{figure}
    \centering
    \includegraphics[width=1.0\linewidth]{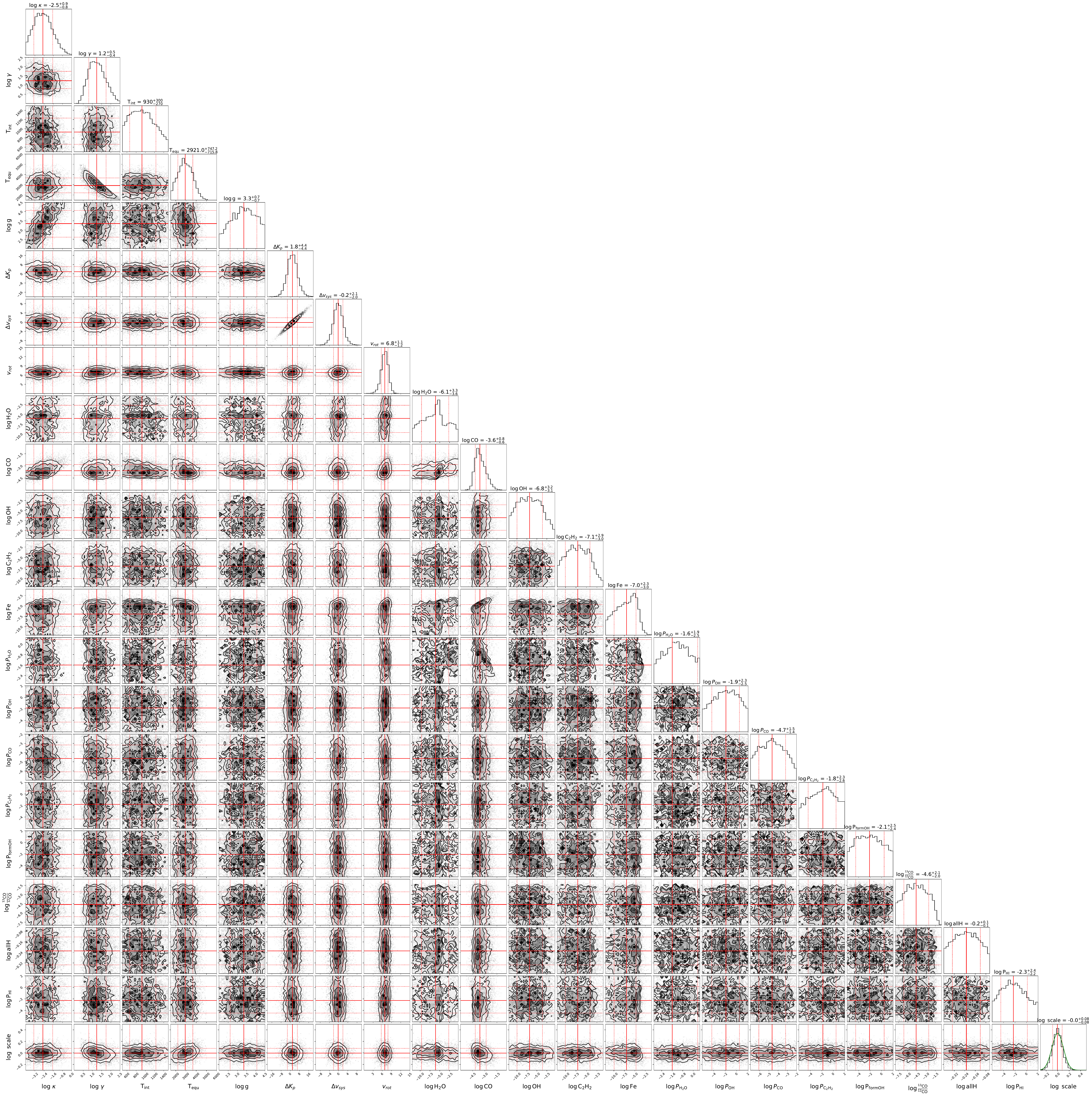}
    \caption{Full corner plot for the retrieval omitting 8 principal components. Red solid lines indicate the medians, while red dashed lines indicate the bounds of the marginalized 68\% confidence interval. We discuss these results in Section \ref{sec:res}.}
    \label{fig:corner8}
\end{figure}

\begin{figure}
    \centering
    \includegraphics[width=1.0\linewidth]{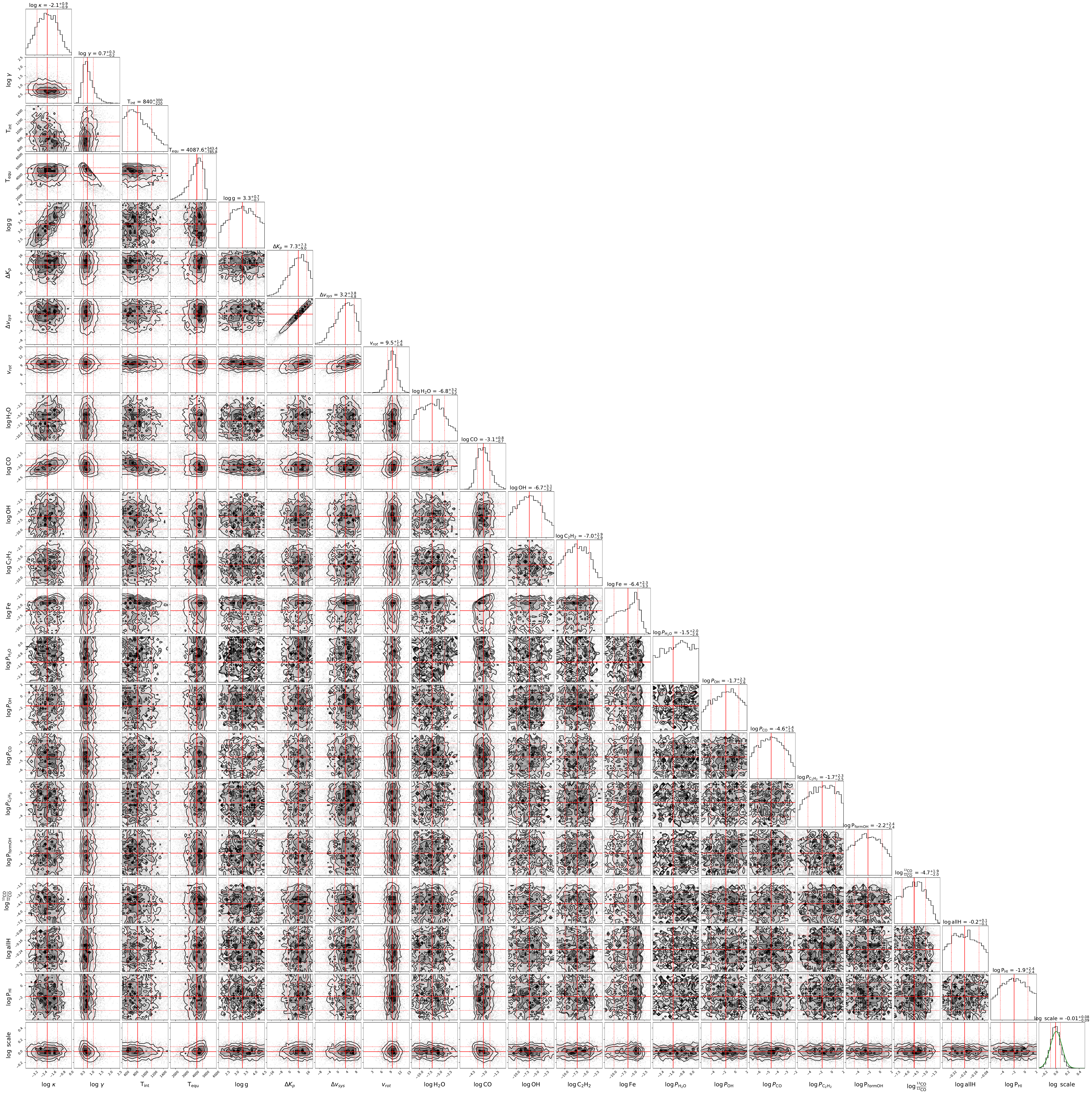}
    \caption{Full corner plot for the retrieval omitting 12 principal components. Red solid lines indicate the medians, while red dashed lines indicate the bounds of the marginalized 68\% confidence interval. We discuss these results in Section \ref{sec:res}.}
    \label{fig:corner12}
\end{figure}

Figures \ref{fig:corner4}, \ref{fig:corner6}, \ref{fig:corner8}, and \ref{fig:corner12} present the full corner plots from the retrievals omitting 4, 6, 8, and 12 principal components respectively. The units, priors, and retrieved quantity are listed in Table \ref{tab:priors}. In cases where parameters are well-constrained, the plotting limits have been adjusted to focus on the posterior rather than spanning the full prior. The truncated normal prior for the scale factor is overplotted in green. All other parameters used a normal prior covering the full range listed in Table \ref{tab:priors}.



\clearpage
\bibliography{exoplanetbib}{}
\bibliographystyle{aasjournal}



\end{CJK*}
\end{document}